\documentstyle[12pt]{article}
\begin{document}
\newcommand{\threej}[4]{\left[ \begin{array}{cc}
                                #1 & #2 \\
								#3 & #4
								\end{array} \right]}
\newcommand{\fourj}[6]{\left( \begin{array}{ccc}
                               #1 & #2 & #3 \\
							   #4 & #5 & #6
							   \end{array} \right)}
\newcommand{\twoj}[4]{\left[ \begin{array}{cc}
                              #1 & #2 \\
							  #3 & #4
							  \end{array} \right]}
\begin{flushright}
IMSc. 93/4 \\
\end{flushright}
\vspace{.3cm}

\begin{center}
{\b {THE REPRESENTATIONS OF TEMPERLEY-LIEB-JONES ALGEBRAS
}}

\vspace{1.5cm}

{\bf R.K. Kaul}$^\star$ \\
{\it The Institute of Mathematical Sciences} \\
{\it Taramani, Madras 600 113, India}

\end{center}

\vspace{2cm}

\baselineskip=24pt

\noindent{\bf Abstract}

Representations of braid group obtained from rational conformal field theories
can be used to obtain explicit representations of Temperley-Lieb-Jones
algebras.
The method is described in detail for SU(2)$_k$ Wess - Zumino conformal field
theories and its generalization to an arbitrary rational conformal field theory
outlined. Explicit definition of an associated linear trace operation in
terms of a certain matrix element in the space of conformal blocks of such
a conformal theory is presented. Further for every primary field of a rational
conformal field theory, there
is a subfactor of hyperfinite II$_1$ factor with trivial relative commutant.
The index of the subfactor is given in terms of identity - identity element of
certain duality matrix for conformal blocks of four-point correlators.  Jones
formula for index ( $<$ 4 ) for subfactors corresponds to spin ${\frac{1}{2}}$
representation of SU(2)$_k$ Wess-Zumino conformal field theory.  Definition of
the trace operation also provides a method of
obtaining link invariants explicitly.

\vspace{2.5cm}

\hrule
\vspace{.3cm}

{\footnotesize $^\star$ email: ~kaul@imsc.ernet.in}

\newpage

\noindent{\bf 1. Introduction}

\vspace{.5cm}

In recent years, a close relationship has emerged between the areas of physics
such as two-dimensional conformal field theory (i.e., critical phenomena in two
dimensions, completely integrable models) and those of mathematics such as
infinite dimensional Lie algebras, von Neumann algebras, braid theory and
topology of three dimensional manifolds with or without knots in them.

One such link between physics and mathematics is provided by topological
quantum field theories.  Indeed such a description of knot theory has proven
powerful
enough to yield many a deep mathematical result$^{1-5}$.  The Chern-Simons
functional integrals over three-manifolds with appropriate Wilson lines
connecting points on the boundaries, provide representations of the groupoid
of braids made of individual strands carrying arbitrary colours and
orientations.  These representations can be used to obtain a whole
variety of invariants for multicoloured links$^{1,3,4}$.  Jones polynomial is
the
simplest invariant in this class, where all the component knots carry one
colour characterized by spin ${\frac{1}{2}}$ representation of $SU(2)$. An
explicit
and complete method of computing expectation values of Wilson link
operators in the $SU(2)$ Chern-Simons theory on $S^3$ has been presented
in refs.(4). This thereby provides a way of obtaining the invariants for
an arbitrary multicoloured link.  Along with the theory of multi-coloured and
oriented braids, two dimensional conformal field theories play an important
role in these developments.

In his original method of obtaining now famous Jones link polynomials$^6$,
Jones
constructed a new representation of the braid group.  This representation is
also related to a representation of a certain finite dimensional von Neumann
(Temperley-Lieb-Jones) algebras which he had come across in his earlier studies
of subfactors of a type
II$_1$ factor.  In these studies, Jones introduced a notion of an index which
in some sense measures the size of a subfactor in a hyperfinite II$_1$
factor.  This index takes values either as $4 cos^2 \frac{\pi}{\ell}, \ell \in
{\bf N}, \ell \ge 3 $ or in [4, ${\infty}$).  In the former case (index ${<
4}$), the
subfactor always has a trivial relative commutant.  This is not so in the case
where the index is greater than four.  All subfactors with index ${\ge 4}$
studied by Jones have non-trivial commutants.  It is of interest to find
possible values of the index on the half-line ${[4, \infty)}$ for the case
where subfactors have trivial relative commutants.  Wenzl has obtained
such values$^7$.  Since Jones work  many a study of subfactors and their index
has
appeared$^{8-11}$.

In this paper, we shall present a general and systematic
method of obtaining the representations of Temperley-Lieb-Jones algebras.
The techniques of conformal
field theory will be  exploited for this purpose.  We shall study one such
theory, namely $SU(2)_k$ Wess-Zumino field theory in detail.  The method
generalizes to any other compact semi-simple group $G$. Further we show that
there
is a subfactor for hyperfinite II$_1$ factor with trivial commutant for every
primary field of the Wess-Zumino level $k$ conformal field theory based on a
compact semi-simple Lie group $G$.  The index of this subfactor is given by the
square of $q$-dimension of the primary field, with $q = exp (2\pi
i/(k+C_V)),C_V$
is the quadratic Casimir for the adjoint representation of the group $G$.
Other
rational conformal field theories, for example the minimal series (central
charge ${C~<~ 1}$) also provide solutions to the problem.  This follows from
an obvious adaptation of the method to such theories.  In fact there is a
subfactor (with trivial relative commutant) for every primary field of any
rational
conformal field theory.  The index for such a subfactor is given in terms of
certain  element of the duality matrix for the four-point correlator of
the primary field.  The matrix element relevant here corresponds to the four-
point conformal blocks where identity operator lives on the internal lines.
The values of index so obtained agree with those in refs. 7 and 11.

The paper is organised as follows :  In section 2, we present the required
aspects of Temperley-Lieb-Jones algebras and discuss the representation
obtained
by Jones and the related representation of the braid group.  In section 3, a
class of representations of braid group will be presented.  These are
developed in the context of level $ k~ SU(2)$ Wess-Zumino conformal field
theories.
These representations then will be used to construct representations of
Temperley-Lieb-Jones algebras along with an explicit presentation of a concept
of
trace in sections 4 and 5.  All these correspond to subfactors of a hyperfinite
II$_1$ factor with trivial commutant.  The generalization of the construction
for a Wess-Zumino conformal field theory based on an arbitrary compact
semi-simple group $G$ will then be discussed only briefly in section 6.  In
fact
the procedure is valid for any rational conformal field theory.  Some
concluding remarks will be presented in the last section 7.  In particular it
will be indicated that the definition of trace developed in section 5 also
yields link invariants.

\vspace{1cm}

\noindent {\bf 2. Temperley-Lieb-Jones algebras}

\vspace{.5cm}

We shall be interested in the von Neumann algebras $A_{m-1}$ generated by
{\bf 1} and projectors $e_1, e_2, \ldots e_{m-1}$ which obey :

$$
\begin{array}{l}
(i)\quad \quad \quad  ~~e^2_i = e_i , \, \quad\quad  e^{\ast}_i = e_i
\\
(ii)\quad\quad\quad  ~e_i \, e_{i\pm 1} \, e_i = \tau e_i
\\
(iii)\quad\quad\quad e_i e_j = e_j e_i \quad\quad\quad  \vert i - j \vert \ge 2
\end{array}    \eqno(2.1)
$$

\noindent These algebras in addition admit of a trace, denoted by "tr"
which is
defined over ${\bigcup }^{\infty}_{m=1} ~A_{m-1}$ and determined
by the normalization tr {\bf 1} = 1 and the following conditions :

$$
\begin{array}{l}

(iv)\quad\quad\quad tr xy = tr yx,\quad \quad x,y \in A_{m-1}
\\
(v)\quad\quad\quad  tr x e_m = \tau tr x, \quad\quad  x \in A_{m-1}
\\
(vi)\quad\quad\quad tr x^\ast y > 0, \quad\quad  x,y \in A_{m-1}
\end{array}
   \eqno(2.1')
$$
Jones has obtained a representation of this algebra and a corresponding
representation of the braid group$^6$.  For a braid generated by
$b_i, ~~i = 1,2, \ldots,$ we have the defining relations :

$$
\begin{array}{lcl}
b_i b_{i+1} b_i & = & b_{i+1} b_i b_{i+1} \\
\\
b_i b_j & = & b_j b_i \quad\quad\quad  \vert i-j \vert \le 2
\end{array}
    \eqno(2.2)
$$
Clearly, if we define

$$
b_i \  = \ q^{\frac{1}{2}} (q-(1+q) e_i)    \eqno(2.3)
$$
with
$$
b^{2}_i - (q^{\frac{3}{2}} - q^{\frac{1}{2}}) b_i - q^2 = 0
     \eqno(2.4)
$$
the $e_i$'s satisfy  relations (2.1) above with $\tau^{-1} =
( q^{\frac{1}{2}} + q^{- \frac{1}{2}})^{2} $.  Eqn.(2.4) corresponds to a
representation of the braid generators where spin ${\frac{1}{2}}$
representation of $SU(2)$ is placed on the strands.  It is this representation
that finally led Jones to his knot invariants.  Further $\tau^{-1}$ is the
Jones index of the subfactor of II$_1$ factor. As stated earlier, Jones theorem
restricts the
values of this index from above representation to $\tau^{-1} =
( q^{\frac{1}{2}} + q^{- \frac{1}{2}})^2$ with $ q = exp {\frac{2\pi i}
{\ell}}, \ell = 3,4, \ldots $.  We shall  later see that these values
correspond to the square of $q$-dimension of spin ${\frac{1}{2}}$
representation and is related to the duality matrix for the correlators of
four spin  ${\frac{1}{2}}$ primary fields in $ SU(2)_k$ Wess-Zumino model
(with the identification ${k = \ell - 2}$).

The algebra (2.1) written in terms of generators~ ${U_i = \tau^{- 1/2} e_i}$
with $\tau^{-1} =$ $ 4 cos^2 {\lambda} $ is known  as Temperley-Lieb algebra in
physics literature, and it first appeared in the context of exactly solvable
statistical mechanical models$^{12}$. We shall refer to the algebras
$(2.1)$ with more general $\tau$ as Temperley-Lieb-Jones algebras.

In the following sections, we shall generalize the relation above between the
representations of the braid group and those of von Neumann algebras (2.1).
Using the von Neumann algebra generated by ${ \{ } e_1, e_2, e_3, \ldots {\}}$
as the hyperfinite II$_1$ factor and that generated by ${\{} e_2, e_3, \ldots
{\}} $ as the subfactor$^{6}$, we shall obtain the index $ \tau^{-1}$ for this
subfactor.

\vspace{1cm}

\noindent{\bf 3. A class of representations for braid generators}

\vspace{.5cm}

Now we shall present the necessary discussion of correlators of level$ ~k~~
SU(2)$
Wess-Zumino conformal field theory on $S^2$.  Their duality properties will be
recapitulated.  The representations of braid group will be given in terms of
the monodromy properties of these correlators$^{13,14,3,4}$, which in turn will
be
used in sections 4 and 5.

$SU(2)_k$ Wess-Zumino conformal field theory admit of ${k+1}$ primary fields
(integrable representations) whose spins are given by ${j = 0, 1/2, 1, \ldots,
k/2}$.  The fusion rules are given by
$$
(j_1) \otimes (j_2) \,=\, ( \vert j_1 - j_2 \vert ) \oplus (\vert j_1 - j_2
\vert + 1) \oplus \ldots \oplus min (j_1 + j_2 ; k-j_1 - j_2)    \eqno(3.1)
$$
\noindent The correlators of this theory can be given in terms of conformal
blocks.  There are more than one complete, but equivalent, sets of conformal
blocks for a given correlator.  These are related to each other by duality
transformations.  For example for the correlator for four primary fields in
representations ${j_1, j_2, j_3, j_4}$, two sets of conformal blocks are
represented by diagrams as shown in fig.1.  In this diagrammatic representation
spins meeting at every trivalent point (e.g. here
${(\ell j_1 j_2), (\ell j_3 j_4), (m j_2 j_3), (m j_1 j_4)}$) satisfy fusion
rules (3.1).  The correlator is non-zero only if the product of the four spins
${j_1, j_2, j_3, j_4}$ contains a singlet.  The duality matrix relating these
two sets of conformal blocks is given$^{ 14,16,3,4}~$ in terms of Racah
coefficients of
$SU(2)_q$ with deformation parameter  related to the level $k$ of the conformal
field theory as ${q = exp 2\pi i/(k+2)}$ :

$$
{a_{\ell m}} \left[ \matrix{ j_1 & j_2 \cr j_3 & j_4} \right] =
(-)^{(j_1 + j_2 + j_3 + j_4)} \sqrt{[2\ell +1][2m+1]} \left( \matrix{
j_1 & j_2 & \ell \cr j_3 & j_4 & m } \right) \eqno(3.2)
$$
$$
\left( \matrix{ j_1 & j_2 & \ell \cr j_3 & j_4 & m} \right) = \Delta
(j_1j_2 \ell) \Delta (j_3j_4\ell) \Delta (j_1j_4,m) \Delta
(j_2 j_3 m) \hspace{3cm}
$$
$$ \begin{array}{ll}
& \hspace{2.7cm} \times \sum_{z \ge 0} (-)^z [z+1]! ~~ {\bf \{ } [z-j_1 - j_2 -
\ell]! \\
&\hspace{3.8cm} \times [z - j_3 - j_4 - \ell]! [z - j_1 - j_4 - m]!  \\
& \hspace{3.8cm} \times [z - j_2 - j_3 - m]! [j_1 + j_2 + j_3 + j_4 - z]! \\
&\hspace{3.8cm}  \times  [j_1+j_3+\ell+m-z]! [j_2+j_4+\ell+m-z]!
 { \bf \} } ^{-1}
\end{array}
$$
where
$$
\Delta (abc) \,=\, \sqrt{\frac{[-a+b+c]! [a-b+c]! [a+b-c]!}
{[a+b+c+1]!}}
$$

\noindent Here $ [a]! = [a] [a-1] \ldots [3] [2] [1] $ and square brackets
represent $q$-numbers :
$$
[x] \, =\, \frac{q^{x/2} - q^{-x/2}}{q^{1/2} - q^{-1/2}}
$$
\noindent These duality matrices satisfy the orthogonality condition :

$$
\sum_{\ell} \, a_{\ell m} \threej {j_1} {j_2} {j_3} {j_4} \, a_{\ell m'} \,
\threej {j_1} {j_2} {j_3} {j_4} \, = \, \delta_{m m'}     \eqno(3.3)
$$

The duality properties of higher point correlators can be constructured in
terms of these duality properties of four-point correlators.  In particular, we
shall be interested in the duality properties of $n$-point correlators as drawn
in figs.2 (a) and (b).  Here we have the same spin $j$ representation on all
the
$n$ external legs.   The $SU(2)$ spins ${(r_1, r_2, \ldots r_{i-1}, r_{i},
r_{i+1}
\ldots r_{n-3})}$ and ${(r_1, r_2, \ldots r_{i-2}, \ell_{i}, r_{i}, r_{i+1}
\ldots r_{n-3})}$ respectively on the interval lines in these figures are in
accordance with the
fusion rules above.  With this $n$-point correlator we associate a Hilbert
space
spanned by orthonormal basis vectors corresponding to these conformal blocks
$ \vert \phi_{r_{1}r_{2}\ldots r_{n-3}} \rangle$ or equivalently
$ \vert \psi^{(i)}_{r_1 r_2 \ldots r_{i-2}(\ell_{i}) r_{i} r_{i+1}
\ldots r_{n-3}} \rangle $ referring to the two diagrams (a) and (b) in fig.2
respectively :
$$
\quad\quad\quad \langle  \phi_{r_{1}r_{2}\ldots r_{n-3}}\, \vert
\phi_{r^\prime_{1}r^\prime_{2}\ldots r^\prime_{n-3}} \rangle \,=\,
\prod_{p}\, \delta_{r_{p} r^\prime_{p}}    \eqno(3.4)
$$
$$
\langle  \psi^{(i)}_{r_1 r_2 \ldots r_{i-2} (\ell_{i}) r_{i},
\ldots r_{n-3}} \, \vert  \psi^{(i)}_{r^\prime_1 r^\prime_2 \ldots
r^\prime_{i-2} (\ell^\prime_{i}) r^\prime_{i} \ldots r^\prime_{n-3}} \rangle
\,=\, \delta_{\ell_{i}\ell^\prime_{i}}\,
\prod_{p}\, \delta_{r_{p} r^\prime_{p}} \quad\quad  \eqno(3.5)
$$

\noindent These two sets of orthonormal vectors are related by the duality
property
$$
\vert  \psi^{(i)}_{r_1 r_2 \ldots r_{i-2} (\ell_{i}) r_{i}
\ldots r_{n-3}} \, \rangle \,=\, \sum_{r_{i-1}}\, a_{r_{i-1} \ell_{i}} \,
\threej {r_{i-2}} {j} {j} {r_{i}} \, \vert \phi_{r_1 r_2 \ldots r_{n-3}} \,
\rangle   \eqno(3.6)
$$

\noindent where $ a_{\ell m} \threej {j_1} {j_2} {j_3} {j_4} $ is the duality
matrix above in (3.2).

These $n$-point conformal blocks provide a basis for a class of matrix
representations of braid group.  Consider a braid of $n$ strands where every
strand carries a spin $j$ of $SU(2)$ and has same orientation.  Such a braid is
generated by generators $b_i,~~ i = 1, 2, \ldots n-1$ of fig.3.  Now the
generator $b_i$ is diagonal in the basis associated with conformal blocks drawn
in fig.2(b) :

$$
b_i \vert  \psi^{(i)}_{r_1 r_2 \ldots r_{i-2} (\ell_{i}) r_{i}
\ldots r_{n-3}} \, \rangle \,=\, \lambda_{\ell_i}\, \vert
\psi^{(i)}_{r_1 r_2 \ldots r_{i-2} (\ell_{i}) r_{i}
\ldots r_{n-3}} \, \rangle \eqno(3.7)
$$

\noindent The eigenvalues are given in terms of the conformal weights of the
representations $ h_{j}  \,=\, \frac{j(j+1)}{k+2} $. Consider a braiding
by one unit of right-handed half-twist in two parallely oriented strands,
both carrying spin $j$ representation. This is is represented by a braid
matrix. Then for every spin
$ \ell = 0,1, \ldots$ min $(2j, k-2j) $ representation in the product
$(j) \otimes (j) \,=\, \oplus (\ell)$ as dictated by fusion rules, we have an
eigenvalue of this braid matrix
given by$^{3,4}$ :

$$
\lambda_{\ell}(j,j) \,=\, (-)^{\ell} \, exp \pi i (4h_j - h_\ell) \,=\,
(-)^{\ell} q^{2j(j+1)-\ell(\ell+1)/2}  \eqno(3.8)
$$

\noindent The braid generators $b_i$ and $b_{i\pm1}$ are diagonal in the bases
$\vert \psi^{(i)} >$ and $ \vert \psi^{(i\pm1)} > $ respectively.  Using
duality properties (3.6), it can easily be seen that these two bases,
$\vert \psi^{(i)} >$ and $ \vert \psi^{(i\pm1)} > $, are related to each other
by
$$
\langle \,  \psi^{(i)}_{r_1 r_2 \ldots r_{i-2} (\ell_{i}) r_{i}
\ldots r_{n-3}} \, \vert \,
\psi^{(i+1)}_{r_1 r_2 \ldots r_{i-1} (\ell_{i+1}) r_{i+1}
\ldots r_{n-3}} \, \rangle \, =\, A_{\ell_i \ell_{i+1}}
$$
$$
A_{\ell_i \ell_{i+1}} \,=\,  a_{r_{i-1}\ell_{i}}\,
\threej {r_{i-2}} {j} {j} {r_i} \, a_{r_{i}\ell_{i+1}} \, \threej
{r_{i-1}} {j} {j} {r_{i+1}}    \eqno(3.9)
$$

Now equations (3.7) - (3.9) define representations of the braid generators.
Using the properties of $q$-Racah coefficients of eqn.(3.2), one can
verify that these  representations do indeed satisfy the defining
relations (2.2) of the braid group.

Alternatively, we may specify these representations of braid generators in
terms of basis $ | \phi > $ associated with conformal blocks of fig.(2a)
instead of the basis $ | \psi > $(fig.2b) above.  Then
$$
\hspace{2cm} b_i \vert \phi_{r_1 r_2 \ldots r_{n-3}} \rangle \,=\,
\sum_{\ell_i, r^\prime_{i-1}} \, \lambda_{\ell_i}\, a_{r^\prime_{i-1}\ell_i}\,
\threej {r_{i-2}} {j} {j} {r_i} \,  a_{r_{i-1}\ell_i}\,
\threej {r_{i-2}} {j} {j} {r_i}
$$
$$
\hspace{4.5cm} \vert
\phi_{r_1 r_2 \ldots r_{i-2} r^\prime_{i-1} r_i \ldots r_{n-3}} \rangle
\eqno(3.10)
$$

The representation of the braid group that Jones used in (2.3) and (2.4) to
relate it to the representation of algebra (2.1) corresponds to spin
$ j = 1/2 $ being placed on every strand of the braid.  For this the
eigenvalues of the braid generators from eqn.(3.8) are :
$ \lambda_{0} = q^{3/2} $ and $ \lambda_{1} = -q^{1/2} $.  Thus, these braid
generators satisfy characteristic relation (2.4).  Corresponding to other
representations ${(j = 1, 3/2, \ldots)}$ of the braid generators above, there
exist new representations of  algebras (2.1).  Now we shall
present these.

\vspace{1cm}

\noindent{\bf 4. Representations of Temperley-Lieb-Jones algebras $ A_{m-1} $}

\vspace{.5cm}

Consider an identity braid of $2m$ strands, each carrying $SU(2)$ spin $j$ with
first
$m$ strands in one orientation and the rest in opposite orientation.  On this
we
apply braid generator $ b_1, b_2, \ldots $ to obtain an arbitrary braid.
Define projectors $ e_i,~ i = 1,2, \ldots m-1 $ as :

$$
e_i \,=\, \frac{(\lambda_{n} - b_{i}) (\lambda_{n-1} - b_{i}) \ldots
(\lambda_{1} - b_i)} {(\lambda_{n} - \lambda_{0}) (\lambda_{n-1} - \lambda_{0})
\ldots (\lambda_{1} - \lambda_{0})}      \eqno(4.1)
$$

\noindent where~~ $ \lambda_{\ell}, ~~\ell = 0,1, \ldots n \equiv$  min$(2j,
k-2j) $
are the eigen values of the braid generators given by (3.8) and hence
$$
\prod^{n}_{\ell =0} \, (\lambda_{\ell} - b_{i}) \, = \, 0
$$

\noindent This implies,
$$
b_i e_i \,=\, e_i b_i \,=\, \lambda_{0} e_{i}      \eqno(4.2)
$$

\noindent and therefore $ e^2_i \,=\, e_i $ which is condition (i) of (2.1).
Notice each $e_i$ has one eigenvalue 1 and $n$ eigenvalues all zero :
$$
e_i \, \vert \psi^{(i)}_{r_1 r_2 \ldots r_{i-2} (\ell_i) r_i \ldots r_{2m-3}}
\, \rangle  \,=\, \delta_{\ell_i 0} \, \vert
\psi^{(i)}_{r_1 r_2 \ldots r_{i-2} (\ell_i) r_i \ldots r_{2m-3}}
\, \rangle     \eqno(4.3)
$$

\noindent This may be equivalently reexpressed in basis $ | \phi > $ as

$$
 e_i \, \vert \phi_{r_1 r_2 \ldots r_{2m-3}} \rangle \, =\,
\sum_{r^\prime_{i-1}} \, a_{r^\prime_{i-1} 0} \, \threej {r_{i-2}} {j} {j}
{r_{i}} \, a_{r_{i-1}0} \,  \threej {r_{i-2}} {j} {j} {r_{i}}~~~
 \vert \phi_{r_1 r_2 \ldots r_{i-2} r^\prime_{i-1} r_i r_{2m-3}} \rangle
\eqno(4.4) $$

\noindent Now since,
$$
a_{\ell 0} \, \threej {j_1} {j_2} {j_3} {j_4} \,=\,(-)^{j_1 +j_3 -\ell}
{\frac{\sqrt {[2\ell+1]}}
{\sqrt {[2j_1 +1] [2j_3+1]}}} \, \delta_{j_1 j_4}\, \delta_{j_2 j_3}
$$

\noindent we have
$$
e_i \, \vert \phi_{r_1 r_2 \ldots r_{2m-3}} \rangle \, =\,
\sum_{r^\prime_{i-1}} \, \delta_{r_i r_{i-2}} \,(-)^{(2j+2r_i -r_{i-1}
-r^{\prime}_{i-1})}
  \frac{\sqrt{[2r_{i-1}+1] [2r^\prime_{i-1}+1]}}{[2j+1] [2r_{i-2}+1]} \,
\vert \phi_{r_1 r_2 \ldots r_{i-2} r^\prime_{i-1} r_i \ldots r_{2m-3}} \rangle
 \eqno(4.5)
$$

Now to check that the generators ${e_i}$ so defined do indeed satisfy condition
(ii) of (2.1), we use (4.4) or (4.5) to calculate $ {e_i e_{i \pm 1} e_i} $
$$
e_i e_{i \pm 1} e_i \,  \vert \phi_{r_1 r_2 \ldots r_{2m-3}} \rangle \, = \,
\left( a_{00} \twoj {j} {j} {j} {j} \right)^2 \, e_i \,  \vert
\phi_{r_1 r_2 \ldots r_{2m-3}} \rangle
$$

\noindent The following identifies are useful in proving this relation :

$$
\sum_{r^\prime_{i-1}} \, a_{r^\prime_{i-1}0} \, \threej {r_{i-2}} {j} {j}
{r_i} \, a_{r_i 0} \, \threej {r^\prime_{i-1}} {j} {j} {r_{i+1}} \hspace{6cm}
$$
$$\hspace{1cm} \times a_{r^\prime_{i}0} \,  \threej {r^\prime_{i-1}} {j} {j}
{r_{i+1}} \,
a_{r^\prime_{i-1}0} \, \threej {r_{i-2}} {j} {j} {r^\prime_i} \,
= \delta_{r_i r^\prime_i} \, \delta_{r_i r_{i-2}} {1 \over {[2j+1]^2} }
$$
$$
\sum_{r^\prime_{i-1}} \, a_{r^\prime_{i-1}0} \, \threej {r_{i-2}} {j} {j}
{r_i} \, a_{r_{i-2} 0} \, \threej {r_{i-3}} {j} {j} {r^\prime_{i-1}}
\hspace{6cm}
$$
$$
\hspace{1cm} \times a_{r^\prime_{i-2} 0} \,  \threej {r_{i-3}} {j} {j}
{r^\prime_{i-1}} \,
a_{r^\prime_{i-1}0} \, \threej {r^\prime_{i-2}} {j} {j} {r_i} \,
=\delta_{r_{i-2} r^\prime_{i-2}} \, \delta_{r_i r_{i-2}} {1 \over{[2j+1]}}
$$

\noindent Now since $ a_{00} { \twoj {j} {j} {j} {j} } \, = \, (-)^{2j}~ \left(
[2j+1]
\right)^{-1} $ we, have

$$
e_i \, e_{i \pm 1} \, e_i \,=\, \frac{1}{[2j+1]^2} \, e_i      \eqno(4.6)
$$

The equations (4.3) or (4.4) or (4.5) define matrix representations of the
Temberley-Lieb-Jones
algebras (2.1) with

$$
\tau (j) \,=\, \left( a_{00} \twoj {j} {j} {j} {j} \right)^2 \, =
\frac{1}{[2j+1]^2}      \eqno(4.7)
$$

\vspace{1cm}

\noindent{\bf  5. Trace}

\vspace{.5cm}

The algebras ${A_{m-1}}$ generated by ${(1, e_1, e_2 \ldots e_{m-1})}$ where
$e_i$'s are as given above and satisfy the conditions (2.1) with
$\tau$ as in eqn.(4.7), also admit of a linear trace.  This trace defined
on $ {\bigcup}^{\infty}_{m=1} ~A_{m-1}$, satisfies the
properties listed in (2.1${^\prime}$).  The structure of the algebras $A_{m-1}$
is essentially
decided by this trace.  We shall now present an explicit definition of this
trace.  To this purpose consider the conformal blocks for $2m$ primary fields,
all in spin $j$ representation, diagrammatically represented by fig.4(a).
We have labeled the first $m$ primary fields by ${1,2,3, \ldots m }$ and the
rest
are denoted by ${m^\prime, (m-1)^\prime, \ldots 2^\prime, 1^\prime}$ as
indicated.  The orthonormal basis vectors of the associated vector space
corresponding to these conformal blocks will be denoted by ${ | \chi_{\ell_1
\ell_2 \ldots \ell_{2m-3}} \rangle}$.  Fig.4(b) is a redrawing
of fig.2(a) with this new convenient labeling of the lines.  This corresponds
to the basis vector ${ | \phi_{r_1 r_2 \ldots r_{m-1} s_{m-2} \ldots s_2 s_1}
\rangle }$ with identification ${s_{m-1} \equiv  r_{m-1}}$.  These two sets of
vectors are related to each other by duality transformation which can be
constructed in terms of the elementary duality transformation involving four
representations at a time (fig.1).  The result of this exercise is :

$$
\vert \chi_{\ell_1 \ell_2 \ldots \ell_{2m-3}} \rangle \,=\, \sum_{r_i s_i}\,
A_{(\ell_1 \ell_2 \ldots \ell_{2m-3}) (r_1 r_2 \ldots r_{m-1} s_{m-2} \ldots
s_2 s_1)}\,  \vert \phi_{r_1 r_2 \ldots r_{m-1} s_{m-2} \ldots s_2 s_1}
\rangle     \eqno(5.1)
$$

\noindent with

$$
A_{(\ell_1 \ell_2 \ldots \ell_{2m-3}) (r_1 r_2 \ldots r_{m-1} s_{m-2} \ldots
s_2 s_1)} \,=\, \prod^{m-1}_{p=1} \, a_{r_p \ell_{2p}} \, \threej {r_{p-1}} {j}
{\ell_{2p-1}} {s_p} \, a_{s_p \ell_{2p+1}} \, \threej {r_{p-1}} {\ell_{2p}}
{j} {s_{p-1}}     \eqno(5.2)
$$

\noindent Here we identify ${ \ell_{2m-2} \equiv j,~ \ell_{2m-1} \equiv 0 ,~~
r_0 \equiv j, ~s_0 = j }$ and ${s_{m-1} \equiv r_{m-1}}$.

Now for ${ x \in A_{m-1}}$ , we define the trace "tr" as the following matrix
element:
$$
tr~ x = \langle \chi_{o j oj \ldots ojo} \vert~ x ~\vert
\chi_{o j oj \ldots ojo} \, \rangle   \eqno(5.3)
$$

\noindent This is the matrix element between vectors~ ${ \vert \chi_{\ell{_1}
\ldots \ell_{2m-3}} \, \rangle }$~ (fig.(4a)) where all the odd-indexed
internal
lines carry the identity representation, ${\ell_{2p+1} = 0}$ and therefore by
fusion rules all the even indexed lines carry spin $j$ representations,
${\ell_{2p} = j}$.

This definition of trace can as well be written in terms of ${\vert \phi > }$
basis above.  For this notice from (5.2)
$$
A_{(o \ell_2 o \ldots o \ell_{2m-4} o) (r_1 r_2 \ldots r_{m-1} s_{m-2} \ldots
s_2 s_1)} \,=\,(-)^{mj -r_{m-1}}~ \frac{\sqrt{[2 r_{m-1}+1]}}{[2j+1]^{m/2}}  \,
\prod^{m-2}_{p=1}  \delta_{\ell_{2p} j} \, \delta_{r_p s_p}     \eqno(5.4)
$$

\noindent Next let us define the abreviated notation for these vectors
corresponding to conformal blocks shown in fig.5 as
$$
\vert \phi^{(m)}_{(r)} \rangle  \equiv \vert \phi^{(m)}_{(r_1 r_2 \ldots
r_{m-1})} \, \rangle \equiv \vert \phi_{r_1 r_2 \ldots r_{m-2} r_{m-1}
r_{m-2} \ldots r_2 r_1} \, \rangle     \eqno(5.5)
$$

\noindent Then the trace (5.3) can be rewritten as

$$
tr ~ x = \frac{1}{[2j+1]^m} \, \sum_{\stackrel{r_1 \ldots r_{m-1}}{r^\prime_1
\ldots r^\prime_{m-1}}} \,~(-)^{(2mj-r_{m-1} -r^{\prime}_{m-1} )}
{\sqrt{[2r_{m-1}+1] [2r^\prime_{m-1}+1]}} \hspace{2cm}
$$
$$
\hspace{4cm} \times \langle  \phi^{(m)}_{(r_1 \ldots r_{m-1})} \vert~ x ~\vert
\phi^{(m)}_{(r^\prime_{1} r^\prime_{2} \ldots r^\prime_{m-1})} \, \rangle
\eqno(5.6)
$$

\noindent The generators ${e_i, ~i=1,2, \ldots m-1}$ act on the vectors (5.5)
as

$$
e_i \, \vert \phi^{(m)}_{(r_1 r_2 \ldots r_{m-1})} \, \rangle \,=\,
\sum_{r^\prime_{i-1}} \, \delta_{r_i r_{i-2}} \,
{}~(-)^{(2j+2r_{i} -r_{i-1} -r^{\prime}_{i-1})} \hspace{5cm}
$$
$$
\hspace{4cm} \times \frac{\sqrt{[2r_{i-1}+1] [2r^\prime_{i-1}+1]}}{[2j+1]
[2r_{i-2}+1]} \,
{}~~\vert \phi^{(m)}_{(r_1 r_2 \ldots r_{i-2} r^\prime_{i-1} r_i \ldots
r_{m-1})} \, \rangle    \eqno(5.7)
$$

Now it is just a straight forward calculation to check that our definition
(5.3) or equivalently (5.6) of trace does indeed satisfy the properties
(2.1$^{\prime}$).  For example notice from (5.6) :

$$
tr ~ 1 \,=\, \frac{1}{[2j+1]^m} \, \sum_{r_1 \ldots r_{m-1}}
\,(-)^{(2mj-2r_{m-1})}~ [2r_{m-1} + 1]
$$

\noindent Since ${(r_{p-1}, r_p, j)}$, $p = 1,2, \ldots m-2 \, $ (with~$
r_o \equiv j$) ~ satisfy fusion rules, we have
$$
\sum_{r_p} \,(-)^{2r_p}~ [2r_p +1] \, = \,(-)^{2j+2r_{p-1}} ~ [2j + 1]
[2r_{p-1} + 1]
$$
\noindent and thus
$$
\sum_{r_1 r_2 \ldots r_{m-1}} \,(-)^{(2mj-2r_{m-1})}~ [2r_{m-1} + 1] \,=\,
[2j+1]^m
$$

\noindent so that tr 1 = 1.  This is also directly obvious from the definition
as in eqn.(5.3).

Further explicit calculations using definition (5.6) of the trace yield
$$
tr ~e_i = \frac{1}{[2j+1]^{i+1}} ~ \sum_{\stackrel{r_1 r_2 \ldots r_i}
{r^\prime_{i-1}}} (-)^{2(i+1)j+2r_i}~ [2r_i +1]  a_{r^\prime_{i-1}0} \threej
{r_{i-2}} {j}
{j} {r_i} a_{r_{i-1} 0} \threej {r_{i-2}} {j} {j} {r_i}       \eqno(5.8)
$$
\noindent and
$$
tr\,  e_i \, e_j \,=\, tr \, e_j \, e_i
$$
\noindent Thus from (5.8), using $r_{-1} =0, r_0 = r^\prime_{0} = j$ and
 $ a_{j0} \left[ \matrix{0 & j \cr j & r_1 } \right] = \delta_{r_1 0} $ :
$$
tr~ e_1 \,=\,\frac{1}{[2j+1]^2}\,=\, \left( a_{00} \twoj {j} {j} {j} {j}
\right)^2
$$

Now we could use inductive argument of Jones to prove that this trace does
{}~satisfy ~other ~the ~conditions ~in~ (2.1$^{\prime}$) ~with ~~$ \tau \,=\,
\left( a_{00} \twoj {j} {j} {j} {j} \right)^2$  = $ [2j+1]^{-2}$.
Alternatively, we could  evaluate $tr~x e_m$ directly. It is convenient to
use the definition (5.3) instead of (5.6) for this purpose. Thus we evaluate
the matrix element $ \langle \chi_{0j0j...j0} \vert~ x~e_m~ \vert \chi_{0j0j...
j0} \rangle $  where the vector $| \chi \rangle$ here corresponds to
$2(m+1)$-point correlator
(fig.4(a) with $2(m+1)$ external legs instead of $2m$). To do so we first
perform a four-point duality transformation involving external lines
$m$ and $m+1$ on this vector so that the transformed vector is an eigen-
function of $e_m$, and then  operate by $e_m$. This leads to a factor
$ a_{00} \left[ \matrix{j & j \cr j & j } \right] $. Now transform back to the
$|\chi_{0j0j..0j0} \rangle $ basis through another four-point duality
transformation . This yields one more factor of $a_{00} \left[ \matrix{j & j
\cr j & j } \right]$. Thus we have the desired result:

$$
tr~x~e_m ~=~ \left( a_{00} \left[\matrix{j & j \cr j & j} \right] \right)^{2}
tr ~x ~ = ~ {1 \over{[2j+1]}^2} ~tr~ x
$$

This completes our explicit construction of a class of representations of the
algebra (2.1) along with the trace operation of (2.1$^{\prime}$) with the
parameter $\tau$ given by the square of the identity-identity matrix element
of the duality matrix for correlators of four spin $j$  primary fields :
$\tau \,=\, \left( a_{00} \twoj {j} {j} {j} {j} \right)^2 \,=\, [2j+1]^{-2}$.
Notice ${[2j+1]}$ is the $q$-dimension of the spin $j$ representation of
$SU(2)_q$.

\vspace{1cm}

\noindent{\bf 6. Other representations}

\vspace{.5cm}

This method of explicit construction of representations of Temperley-Lieb-Jones
algebras
(2.1) using level $~k~ SU(2)$ Wess-Zumino conformal field theory has an obvious
generalization to the case of Wess-Zumino model based on an arbitrary compact
semi-simple Lie group $G$.  Let $\lambda$ be the weights of a representation
$R$ of
$G$ and $\theta$ the highest root of the Lie algebra normalized as
${(\theta, \theta) = 2}$.  Then the primary fields of $G_k$ Wess-Zumino model
are all those representations with ${(\lambda, \theta) \, \le k}$.  For example
for SU(N)$_k$, ${(\lambda, \theta)}$ being the number of boxes in the first row
of the Young tableaux, the primary fields (integerable representations) are all
those  with number of boxes in the first row of their Young
tableau less than or equal to $k$.  The fusion rules for these representations
are given in terms of their depths.  For every weight $\lambda$ in the highest
weight representation $L{(\Lambda)}$,~ depth ${(\lambda)}$ is the biggest
integer $j$ for which ${\lambda - j \theta \in L(\Lambda)}$.  Then fusion rules
are given$^{11}$ by ${R_1 \, \otimes \, R_2 \,=\, \oplus \, R(\lambda)}$, if
the Clebsch-Gordon coefficient ${C^{R(\lambda)}_{R_1R_2} \, \neq 0}$ and also
if depth $R_1$ + depth $ R_2 \le k - (\lambda, \theta)$.

In order to construct representations of (2.1), we consider a $2m$-strand braid
where we put representation $ R$ (allowed integerable representation of $G_k$
conformal field theory) and its conjugate ${\bar{R}}$ alternately on the $2m$
strands.  Further let us put same orientations on  the first $m$ strands
and opposite orientation on the other $m$ strands. The eigen values
of the braid generators ~$b_{\ell}$~(fig.3)~ are again given in terms of
conformal weights of the
involved representations, ${h_R\,=\, C_R/(C_V+k)}$, where $C_R$ is the
quadratic Casimir of representation $R$ and ${C_V}$ that for the adjoint
representation.  These eigenvalues for braid generators introducing right
handed half twists in two parallely oriented strands carrying representations
$R$ and $\bar{R}$ are given by $^3$.
$$
\lambda_{R_s}(R, \bar{R})\,=\, (-)^{\in_s} \, exp \pi i \,( 4h_R -  h_{R_s}) \,
\equiv
(-)^{\in_s} \, q^{ 2C_R -C_{R_{s}}/2}    \eqno(6.1)
$$

\noindent corresponding to the irreducible representations $R_s$ consistent
with
the fusion rules given above in the product ${R \otimes \bar{R} \,=\,
\oplus^{n}_{s=0} R_s}$  with the identification that $R_0 $ is the identity
representation.  Here
${(-)^{\in_s} \,=\,(-)^{(\lambda_{R_s} , \theta) / 2}  \,=\, \pm 1}$
.The
parameter $q$ is related to the level $k$ of the Wess-Zumino conformal field
theory
through
$$
q \,=\, exp \, \frac{2 \pi i}{k+C_V}       \eqno(6.2)
$$

The braid generators $b_i$ act on the vectors corresponding to $2m$-point
conformal blocks, introducing half-twist in the $i$th and $(i+1)$th strands. We
define a braiding operation $\hat{b}_i~=~ \sigma_i b_i $
, which introduces a
half-twist followed by interchange (${\sigma}_i $ ) of $i$th and $(i+1)$th
labels ( representations
) of the conformal blocks it acts on. The generators ${e_i, ~i = 1,2, \ldots
m-1}$ of the Temberley-Lieb-Jones algebra (2.1)
now can be constructed as in (4.1) in terms of the first $m-1$  braid
generators
$\hat{b}_i$ as :

$$
e_i \,=\, \frac{(\lambda_{R_n} - \hat{b}_i~)~ (\lambda_{R_{n-1}} - \hat{b}_i)~
\ldots \,
{}~~(\lambda_{R_1} - \hat{b}_i)} {(\lambda_{R_n} - \lambda_{R_0})
(\lambda_{R_{n-1}} -
\lambda_{R_0}) \ldots (\lambda_{R_1} - \lambda_{R_0})}    \eqno(6.3)
$$

The definition of trace (5.3) for $SU(2)_k$ also generalizes as the
matrix element between states corresponding conformal blocks of the type shown
in fig.4(a), now  with representations $R$ and ${\bar{R}}$ alternately placed
on
the external legs and all the odd indexed internal lines carrying the identity
representation ${R_0}$.  The generators (6.3) with this definition of trace
then satisfy all the conditions (2.1) and (2.1$^\prime$) with the index $\tau$
given as the square of identity - identity duality matrix element for the four
point correlator for representations ${R, \bar{R}, R, \bar{R}}$ as shown in
fig.6.  This duality matrix ${a_{R_s R_s'} \twoj {R} {\bar{R}} {R} {\bar{R}}}$
here is again given in terms of $q$-Racah coefficients of quantum group $G_q$
with deformation $q$ as given in (6.2).  Further identity-identity element of
this duality matrix is inverse of the $q$-dimension of representation $R$.
Thus
we have
$$
\tau(R) \,=\, \left( a_{R_0 R_0} \twoj  {R} {\bar{R}} {R} {\bar{R}} \right)^2
\,=\, (dim_q R)^{-2}     \eqno(6.4)
$$
Now the $q$-dimension of a representation $R$ of $G_q$ is given by the
"Weyl dimensionality formula":
$$
dim_q R(\lambda) \ = \ \prod_{ \alpha_{+} \in \Delta_{+}} ~\frac{[
(\lambda ,\alpha_{+} ) + ( \delta , \alpha_{+})]}{ [ (\delta , \alpha_{+})]}
$$
where $\lambda$ is the highest weight of the representation $R(\lambda)$ ,
$\Delta_{+}$
is the set of positive roots $ \{ \alpha_{+} \} $ and
$ \delta = {1/2}~ \sum_{\alpha_{+} \in \Delta_{+}} ~\alpha_{+} $ and square
brackets indicate $q$-numbers with $q$ given by eqn.(6.2).

The parameter ${\tau^{-1}}$ is the index (dimension) of the subfactors of
hyperfinite II$_1$ factors with trivial relative commutant.  For $SU(N), ~N \ge
2$,
the formula (6.4) agrees with that obtained by Wenzl$^7$ :

$$
\tau^{-1} \,=\, \prod_{1 \le r < s \le N} ~~ \frac{sin^2 ( (\lambda_r -
\lambda_s + s - r) \pi/\ell)}{sin^2 ( (s-r) \pi/\ell)}    \eqno(6.5)
$$

\noindent where ${\lambda_1, \lambda_2 \ldots \lambda_{N-1}}$ are the
number of boxes in the first, second, third, .... $(N-1)$th rows
respectively in the Young  tableaux
(with $~~ \lambda_1 \le \ell - N$)  of a representation of $SU(N)$.  Here
${\lambda_N}$ is set to zero.  The expression (6.5) is indeed square of the
$q$-dimension of the representation, with ${q = exp \left( \frac{2 \pi i}{k+N}
\right)}$ and identification ${\ell = k + N}$.  Formula (4.7) is the special
case of (6.5) for $N = 2$.  And formula (6.4) generalizes this index of
subfactors of II$_{1}$ factors with trivial commutants for any arbitrary
representation of a compact semi-simple Lie group.

The method of construction of representations of algebra (2.1) presented here
in fact is valid not only for Wess-Zumino conformal field theories but also for
any arbitrary rational conformal field theory;  in particular, for example,
for the minimal series.  Any rational conformal field theory can be
constructed$^{15}$  as a coset construction $G/H$ of Wess-Zumino conformal
field
theories, the Wess -Zumino theories discussed above corresponding to $H$ being
trivial.  Using the properties of conformal blocks for correlators much like
as we have done in section 3 for $SU(2)_k$ Wess-Zumino model, representations
of braid group corresponding to every primary field of a rational conformal
field theory can be constructed.  The relevant braids for our purpose are
those which are obtained from identity braid with primary fields ${\phi}$ and
conjugate field ${\bar{\phi}}$ live
alternately on $2m$ strands. The first $m$ strands are all oriented
in one direction and the other $m$ in opposite direction. The eigen values of
the generators of such
braids are given again in terms of the conformal weights in a similar fashion
as in eqn.(6.1).  The generator ${e_i}$ of the von Neumann algebra (2.1) are
given by a formula of the type (6.3).  Generalization of the trace "tr" of
eqn.(5.3) is also possible.  It is given in terms of the matrix element
between  the states depicted by fig.4(a) where now the external lines
alternately
carry
primary fields $\phi$ and $\bar{\phi}$ of the conformal field theory under
consideration and all the
odd indexed internal lines carry the identity representation $\phi_{0}$.  The
index
${\tau^{-1}}$ is then again related to identity-identity
{}~${(\phi_0-\phi_0)}$~ element of the duality matrix,~
${a_{\phi_0 \phi_0} \twoj {\phi} {\bar{\phi}}  {\phi} {\bar{\phi}}}$ for four
point correlators for primary fields ${{\phi},~ {\bar{\phi}},~ {\phi},~
{\bar{\phi}}}$ as shown by a diagram of the type shown in fig.6.  Thus we have
a general result :

\vspace{.5cm}

{\it For every primary field ${\phi}$ of a rational conformal field theory,
there is a
subfactor of hyperfinite II$_{1}$ factor with ~trivial relative ~commutant with
index
${\left( \tau ( \phi) \right)^{-1}}$, where}

$$
\tau (\phi)\,=\, \left( a_{\phi_0 \phi_0} \,  \twoj {\phi} {\bar{\phi}}
{\phi} {\bar{\phi}} \right)^2     \eqno(6.6)
$$

\vspace{.5cm}

On the other hand the calculations of ref.(11) give this index as
$(S_{\phi_0 \phi}/S_{\phi_0 \phi_0})^{2}$ where ${S_{\phi \phi^\prime}}$
is the $S$-modular transformation ($\tau \rightarrow -1/\tau~$ , where this $\,
\tau$  is the modular parameter of a torus) matrix on the characters
${\chi_{\phi}}$ of
the conformal field theory. Thus our result is consistent with this result with
the identification:

$$
\left( a_{\phi_0 \phi_0} \, \twoj {\phi} {\bar{\phi}} {\phi} {\bar{\phi}}
\right)^2 \, =\, \left( S_{\phi_0 \phi_0} \over {S_{\phi \phi_0}} \right)^2
\eqno(6.7)
$$
These values of index so obtained also agree with those of ref.7.

Now let us study  the minimal models with central charge ${c < 1}$ as other
examples.  These
{}~models ~have a ~coset ~construction$^{15}$ ~based on~~ $SU(2)_m $ $ \otimes
SU(1)/SU(2)_{m+1}$.  The central charge is given by

$$
c \,=\, 1 - \, \frac{6}{(m+2) (m+3)} \quad \quad m = 1,2, \ldots
$$

\noindent and the ~conformal weights for the primary fields~ ${\phi_{(r,s)}}$
$~(1 \le r \le m+1,$~~~  $1 \le s \le m+2)$ are given by

$$
h_{(r,s)} \,=\, \frac{\left\{ (m+3)r - (m+2)s \right\}^2 - 1}{4(m+2) (m+3)}
$$

\noindent The identity operator is ~~${\phi_{(1,1)}}$.  Due to the
{}~identification
{}~~$\phi_{(r,s)} \, \equiv $ $\phi_{(m+2-r, m+3-s)}$ , we pick up the primary
fields
from the grid ${\phi_{(r,s)}}$ by taking $r+s$ to be even.  Now consider the
duality matrix corresponding to fig.1, where the external lines carry the
primary
fields ${\phi_{(r_1, s_1)}, \phi_{(r_2, s_2)}, \phi_{(r_3, s_3)}}$ and
${\phi_{(r_4, s_4)}}$.  The internal lines carry the
representations ${\phi_{(r,s)}}$ and  ${\phi_{(r',s')}}$ respectively in the
two diagrams of this figure.  This  duality matrix is given by the product
of duality matrices for $SU(2)_m$ and $SU(2)_{m+1}$ theories :
$$
a_{\phi_{(r,s)} \phi_{(r',s')}}  \threej {\phi_{(r_1,s_1)}}
{\phi_{(r_2,s_2)}} {\phi_{(r_3,s_3)}} {\phi_{(r_4,s_4)}} \hspace{6cm}
$$
$$
\hspace{3cm} = \,
a^{(1)}_{{\frac{r-1}{2}}{\frac{r'-1}{2}}}
\threej {\frac{r_1 -1}{2}} {\frac{r_2 - 1}{2}} {\frac{r_3 -1}{2}}
{\frac{r_4 - 1}{2}} \quad   a^{(2)}_{{\frac{s-1}{2}}{\frac{s'-1}{2}}}
\threej {\frac{s_1 -1}{2}} {\frac{s_2 - 1}{2}} {\frac{s_3 -1}{2}}
{\frac{s_4 - 1}{2}}
$$

\noindent The duality matrices ${a^{(1)}_{\ell m}, ~~a^{(2)}_{\ell m}}$ on the
right hand side are the same as (3.2) except for that the $q$-parameters are
now
given by ${q_1 \,=\, exp \frac{2\pi i}{m+2}}$ and  ${q_2 \,=\, exp
\frac{2\pi i}{m+3}}$ respectively for these two.

Thus then for every representation ${\phi_{(r,s)}}$ we have a subfactor whose
index can be written as
$$
 \left(\tau (\phi_{(r,s)}) \right)^{-1} \,=\, \left( a_{\phi_{(1,1)}
\phi_{(1,1)}} \,
\threej {\phi_{(r,s)}} {\phi_{(r,s)}} {\phi_{(r,s)}} {\phi_{(r,s)}}
\right)^{-2} \, = \, \left( {[r]_1 [s]_2} \right)^2     \eqno(6.8)
$$

\noindent where square brackets with subscripts 1,2 represent $q$-numbers with
parameters  given respectively by ${q_1}$  and ${q_2}$ above.

We can go through a similar discussion with respect to other discrete series
such as superconformal theories, which have a coset construction based on
${SU(2)_m \times SU(2)_2 / SU(2)_{m+2}}$ or even more general case
based on the coset ${SU(2)_m \times SU(2)_{\ell} / SU(2)_{m+\ell}}$.

\vspace{1cm}

\noindent{\bf 7. Concluding remarks}

\vspace{.4cm}

We have exploited the properties of rational conformal field theories to obtain
representations of Temperley-Lieb-Jones  algebras(2.1).  Correlator conformal
blocks of
these theories provide a basis for developing a class of representation of the
braid group.  These in turn yield representations of the algebras (2.1)
through (6.3) and its generalization to arbitrary rational conformal field
theory.  The eigen value of the braid generator in the direction of identity
operator (for definiteness ${\lambda_{R_{0}}}$ in eqn.(6.3))  plays a special
role
in this construction.  Replacing this eigen value by any other eigen value
does not provide a representation of the algebra (2.1).  Further a trace with
requiste properties (2.1$^{\prime}$) is given in terms of a certain specific
matrix element in the space of conformal blocks.  Thus we have
demonstrated that for every primary field $\phi$ of an arbitrary rational
conformal field theory on ${S^2}$, there is a subfactor of II$_{1}$ factor with
trivial relative commutant.  The index of this subfactor is given by inverse
of the square of identity-identity element of the four point duality matrix
for primary fields $\phi,~{\bar{\phi}},~ \phi, ~{\bar{\phi}}$.  The discrete
series with index ${4 cos^2 ~(\frac{\pi}{\ell}), \,~ \ell = 3,4, \ldots }$
obtained
by Jones corresponds to the spin 1/2 representation of $SU(2)_k$ conformal
field theory with the identification ${\ell = k+2}$.

The explicit definition of trace as developed in section 5 for $SU(2)_k$
conformal field theory and its obvious generalization for an arbitrary rational
conformal field theory can also be used to obtain link invariants.  For this
purpose consider a link represented as the closure of an $m$-strand braid
${\cal B}_m$ written
as a word in terms of generators ${b_1, b_2 \ldots b_{m-1}}$.  We think of this
closure  in terms of $2m$-strand braid ${\hat{\cal {B}}}_{m}$ where all the
braiding (i.e. ${\cal B}_m$) lives in first $m$  parallely oriented strands
, all carrying representation $\phi$ of the conformal field theory. The
rest of the $m$ strands have no braiding but are oppositely oriented and all
carry conjugate representation $\bar{\phi}$. Then
closure of the braid ${\cal B}_m$ constitutes in connecting the first strand
to the last,  the second to the second last and so on from above and below in
 the so constructed $2m$-strand braid ${\hat{\cal B}}_{ m}$.   The invariant
for such a link is given by
$$
V[L] \,=\, \left( a_{\phi_0 \phi_0} \left[ \matrix{ \phi & \bar{\phi} \cr
\phi & \bar{\phi} } \right] \right)^{-m} ~~ tr~ {\hat{\cal B}}_{m}
\eqno(7.1)
$$

\noindent The trace here is the matrix element between vectors representing
conformal blocks for correlators of $2m$ primary fields, first $m$ being all
$\phi$ and the other
$m$ all conjugate $\bar{\phi}$ (of the type as shown in fig.4(a)).
All the odd indexed internal lines are supposed to carry identity represntaion.
Eqn.(7.1) represents  a link invariant because as can be shown this trace
has the property tr ~${\hat{\cal B}}_{m} \, b_m^{\pm 1} = a_{\phi_0 \phi_0}
\left[ \matrix {\phi & \bar{\phi} \cr \phi & \bar{\phi} } \right]~ tr~
{\hat{\cal B}}_{m}$ ~where ~${\hat{\cal B}}_m ~b_m$ ~is a $2(m+1)$ strand braid
with weaving pattern in the first $m+1$ strands. This property
ensures that $ V[L]$ above remains unaltered under Markov moves. Notice
invariant for an unknot is $\left( a_{\phi_0 \phi_0} \left[ \matrix{ \phi &
\bar{\phi} \cr \phi & \bar{\phi}} \right]  \right)
^{-1}$.
The concept of trace used above in (7.1) provides an explicit
presentation of the formal Markov trace  used in
ref(11).

The statement (7.1) gives us link invariant for monocromatic links.  This is a
version of Theorem 6 of second of refs.(4) which gives the link invariant for a
multicoloured link obtained as closure of a braid in an $SU(2)$ Chern-Simons
theory.

We shall discuss the link invariants obtained through (7.1) and their
generalization for multi-colour links for the minimal series of conformal field
theories as well as for super conformal field theories in more detail else
where$^{18}$.

\newpage

\noindent{\bf References}

\vspace{.4cm}

\begin{enumerate}

\item E. Witten : Quantum field theory and Jones polynomials, Commun. Math.
Phys.
{\bf 121} (1989) 351-399.
\item M. Atiyah : {\it The geometry and Physics of knots} Cambridge Univ. Press
(1989).
\item R.K. Kaul and T.R. Govindarajan : Three dimensional Chern-Simons theory
as a theory of knots and links I, Nucl. Phys. {\bf B380} (1992) 293-333 ; \\
Three dimensional Chern-Simons theory as a theory of knots and links II :
multicoloured links, Nucl.Phys.{\bf B393} (1993) 392-412; \\
P. Rama Devi, T.R. Govindarajan and R.K. Kaul: Three dimensional Chern - Simons
theory as a theory of knots and links III : an arbitrary compact semi-simple
group, Nucl. Phys. {\bf B402} (1993) 548-566.
\item R.K. Kaul:  Complete solution of $SU(2)$ Chern-Simons theory, preprint
IMSc./92/56 \\
R.K. Kaul : Chern-Simons theory, coloured-oriented braids and link
invariants, preprint IMSc./93/3, Commun. Math. Phys.(in press).
\item J.M.F. Labastida and A.V. Ramallo : Operator formalism for Chern-Simons
theories, Phys. Lett. {\bf B 227}(1989) 92-102 ; \\
J.M.F. Labastida, P.M. Llatas and A.V. Ramallo : Knot operators in Chern-Simons
gauge theory, Nucl. Phys.  {\bf B348} (1991) 651-692 ; \\
J.M. Isidro, J.M.F. Labastida and A.V. Ramallo : Polynomials for torus links
from Chern-Simons gauge theories, preprint US-FT-9/92.
\item V.F.R. Jones : Index for subfactors, Invent.Math. {\bf 72}(1983) 1-25; A
polynomial invariant for knots via von Neumann algebras, Bull. Am. Math.Soc.
{\bf 12} (1985) 103 - 111; Hecke algebra representations of braid group and
link polynomials, Ann.Math. {\bf 126} (1987) 335-388.
\item H. Wenzl : Hecke algebras of type $A_n$ and subfactors, Invant.Math.
{\bf 92} (1988) 349-383; Quantum group and subfactors of type B, C and D,
Commun. Mat. Phys. {\bf 133} (1990) 383-432.
\item M. Pimsner and S. Popa : Entropy and index for subfactors, Ann. Scient.
Ec. Norm.Sup. {\bf 19} (1986) 57-106.
\item R. Longo : Index of subfactors and statistics of quantum fields I and II,
Commun. Math. Phys. {\bf 126} (1989) 217-247 and Commun. Math. Phys. {\bf 130}
(1990) 285-309.
\item U. Krishnan, V.S. Sunder and C. Varughese: On some subfactors of
integer index arising from vertex models, ISI-Bangalore preprint(1993).
\item J. de Boer and J. Goerre: Markov traces and II$_1$ factors in conformal
field theory, Commun. Math. Phys. {\bf 139} (1991) 267-304.
\item H.N.V. Temperley and E.H. Lieb : Relation between the percolation and
colouring problem, Proc. R. Soc. London {\bf A 322} (1971) 251-280 .\\
See also R.J. Baxter: {\it Exactly solved models in statistical mechanics},
Academic
Press, London (1982).
\item D. Gepner and E. Witten : String theory on group manifold, Nucl.Phys.
{\bf B 278} (1986) 493-549.
\item For general reviews of conformal field theory see :  L. Alvarez-Gaume and
G. Sierra, Topics in conformal field theory, preprint CERN-TH.5540/89  and
P. Ginsparg: Applied conformal field theory, in {\it Fields, strings and
critical phenomenon} ed. E. Brezin and J. Zinn-Justin, North Hollland,
Amsterdam, 1990.
\item A. Tsuchiya and Y. Kanie: Vertex operators in conformal field theory on
$P^1$ and monodromy representations of the braid group, Adv. Studies in
Pure Math. {\bf 16} (1988) 297-372.
G. Moore and N.Seiberg: Classical and quantum conformal field theory,
Commun. Math. Phys. {\bf 123} (1989) 177-254.
\item L. Alvarez Gaume, C. Gomez and G. Sierra : Quantum group interpretation
of
some conformal field theories,Phys. Letts. {\bf 220 B}
(1989) 142-152.
\item P. Goddard, A. Kent and D. Olive : Virasoro algebras and coset space
models, Phys.Lett. {\bf B 152} (1985) 88 and Unitary representations of the
Virasoro and super-virasoro algebras, Commun.Math.Phys. 103 (1986) 105.
\item T.R. Govindarajan, R.K. Kaul and P. Rama Devi, under preparation.
\end{enumerate}

\newpage

\noindent{\bf Figure captions}

\begin{description}

\item[Fig.1.]  Duality of four-point correlators in $SU(2)_k$ Wess-Zumino
theory.

\item[Fig.2.]  Two sets of conformal blocks for correlator of $n$ spin $j$
primary fields.

\item[Fig.3.]  Generator of $n$-braids.

\item[Fig.4.]  Conformal blocks for $2m$-point correlators.

\item[Fig.5.]  Conformal block relevant for the definition of trace.

\item[Fig. 6.]  Duality transformation for four-point correlator in $G_k$
Wess-Zumino conformal field theory.

\end{description}

\end{document}